\documentclass[twocolumn,prb,aps,showpacs]{revtex4}

\usepackage{graphicx}
\usepackage{dcolumn}

\begin{document}

\title{Cooperative domain type interlayer $sp^3$-bond formation in graphite}
\author{Keita Nishioka}
\author{Keiichiro Nasu}
\address{Solid State Theory Division, Institute of Materials Structure Science, High Energy Accelerator Research Organization (KEK), 1-1, Oho, Tsukuba, Ibaraki, 305-0801, Japan}
\date{\today}

\begin{abstract}
Using the classical molecular dynamics and the semiempirical Brenner's potential, we theoretically study the interlayer $\sigma$ bond formation, as cooperative and nonlinear phenomena induced by visible light excitations of a graphite crystal. We have found several cases, wherein the excitations of certain lattice sites result in new interlayer bonds even at non-excited sites. We have also found that, a new interlayer bond is easier to be formed around a bond, if it is already existing. As many more sites are going to be excited, the number of interlayer bonds increases nonlinearly with the number of excited sites. This nonlinearity shows 1.7 power of the total number of excited sites, corresponding to about three- or four-photon process.
\end{abstract}

\pacs{64.70.Nd,61.48.Gh,31.15.xv,05.40.-a}

\maketitle

\section{introduction}\label{sec:introduction}
Recently, the phase transition from the $sp^2$-bonded graphite to the $sp^3$-bonded diamond induced by visible light irradiation has been experimentally proposed, and has attracted considerable attention.\cite{Kanasaki} As well known, the diamond is synthesized macroscopically from the graphite under high temperature and pressure (3000 $^\circ$C, 15 GPa),\cite{Bundy,Irifune} or the irradiation of strong x ray.\cite{Banhart,Nakayama}, due to the high energy barrier between them. It is estimated to be about 0.3-0.4 eV/atom by the local-density approximation (LDA).\cite{Fahy1,Fahy2} However, Kanasaki {\it et al.} found that, after the irradiation of visible photons onto graphite, a new nanoscale domain with $sp^3$ structure, including more than 1000 carbons, appears locally in the graphite surface and stable for more than 10 days at room temperature.\cite{Kanasaki,Radosinski1} The nanoscale domain is an intermediate state between the graphite and the diamond. Therefore, it is called ``diaphite.\cite{Kanasaki,Nature}'' Further pulse excitations will generate a lot of such domains, and they combine with each other, proliferate, and are expected to result in the macroscopic diamond phase finally. In general, this kind of process is called the photoinduced structural phase transition.\cite{Nasu}

The essential aspects of this experiment are as follows: (1) the exciting laser, with an energy of 1.57 eV, should be polarized perpendicular to the graphite layer, while the one polarized parallel to the layer gives no effect. It means that only the interlayer charge-transfer excitation can trigger this process. (2) The exciting light should be a femtosecond pulse, while the picosecond one gives almost no contribution. It means that only a transient generation of an excited electronic wave packet in the semimetallic continuum can efficiently trigger this process. (3) The process is quite nonlinear but less than the ten-photon process.

From the above experiments, we can think of the following scenario for the present process:\cite{Radosinski1} when the femtosecond pulse is irradiated onto the graphite perpendicularly, an electron-hole pair spanning two layers is generated. This electron-hole pair mainly dissipates into the  semimetallic continuum of the graphite as plus and minus carriers due to the good conductivity of graphite. However, by a small but finite probability, this electron-hole pair is expected to be bounded with each other through the interlayer Coulomb attraction. This exciton-like state self-localizes at a certain point of the layer by contracting the interlayer distance only around it. As the local contraction of the interlayer distance proceeds, an interlayer $\sigma$ bond is formed. Through further pulse excitations, a lot of interlayer $\sigma$ bonds are formed stepwise, and then the diaphite structure is expected to appear macroscopically.

In order to clarify the mechanism of this photoinduced phase transition from the graphite to the diaphite, we have already calculated the adiabatic path from the graphite to the small diaphite domain, by means of the LDA (Ref. 11) and also by the semiempirical Brenner's theory.\cite{Ohnishi2,Radosinski2} These results have already well clarified the adiabatic property of the phase transition. We have also studied the early stage dynamics of the interlayer $\sigma$ bond formation along with the above scenario.\cite{Nishioka,Radosinski2} In the previous paper,\cite{Nishioka} by solving the quantum time development of an electron-hole-phonon system, we have described the self-localization of an photoexcited electron-hole pair spanning two layers, full quantum mechanically. Since the subsequent interlayer $\sigma$ bond formation after the self-localization can be taken into account in a classical picture due to the translational symmetry breaking of the excited state,\cite{Nishioka} we can calculate the bond formation using classical molecular dynamics (MD) with the Brenner's potential. Consequently, we have found that the self-localization occurs by the probability of about 2\% when the electron-hole pair is excited as a transient state with the energy of 3.3 $\pm$ 1.8 eV, while the subsequent bond formation is achieved when the excitation energy is more than 4.5 eV, corresponding to about three visible photons with the energy of 1.57 eV.

So far, we have already clarified the dynamics how an interlayer bond is formed. The next stage is to investigate how a lot of interlayer bonds are cooperatively formed by further pulse excitations, that is, the effect of multisite excitation. In the case where several sites in graphite layers are excited, various cooperative phenomena will be expected to influence the interlayer bond formation. To confirm it, we calculate the dynamics of the multisite excitation.

In the starting point of our calculation, we assume that the self-localization has already completed at more than one site in the graphite layers by the irradiation of many visible photons. Hence, in this paper, all the calculations of the bond formation are performed by using the MD, just same as in the previous paper.\cite{Nishioka} Similarly, we adopt the Brenner's potential I in its original version,\cite{Brenner} which can well describe various carbon cluster systems.

\section{model and method}\label{sec:model_method}
We consider only two graphite layers with $AB$ stacking and assume that the interlayer distance contracts locally at several sites of the crystal due to the self-localization of electron-hole pairs. Then, the two opposite carbons at each excited site are intruded inside of the two layers and have a velocity toward the inside. As an initial condition, therefore, we give the carbon pair initial intrusion (potential energy) and an initial kinetic energy. For simplicity, we also assume that the initial displacements and the initial velocities of the two carbons of the pair have the same magnitude in opposite directions. The initial kinetic energy $K_0$ is given by $K_0=2\frac{1}{2}m_cv_0^2$, where $m_c$ is the mass of a carbon atom and $v_0$ is the initial velocity. The initial potential energy is obtained from the relation between the intrusion and the potential energy, calculated by using the Brenner's theory.\cite{Nishioka} In our calculation, the total energy is always conserved, since we do not take into account any energy dissipation effects. The system consists of two graphite layers with the initial interlayer distance 3.35 \AA, and each layer includes 6240 carbons in about 128 \AA\, $\times$ 128 \AA, wherein a periodic boundary condition is imposed.

First, we perform the MD calculation in the case where two or more carbon pairs are excited simultaneously in order to confirm the cooperative phenomena between excitations. In this calculation, we give the same potential energy and kinetic energy to all the excited carbon pairs, as an initial state. We calculate the dynamics for various spatial arrangements, various numbers of excited sites, and various excitation energies.

Second, we investigate the easiness of interlayer bonding when an old interlayer bond has already existed nearby. As an initial state, we give excitation energy to a carbon pair near a certain interlayer bond, which is photogenerated beforehand. In this sense, the situation is a kind of multisite excitation, although the time interval between these two excitations is infinite. We perform the MD calculation for various values of excitation energy and various positions of the excited carbon pairs. Thereby, we can estimate the lowest energy necessary to form an interlayer bond as a function of the distance between the two excited sites.

Finally, we consider the many and random multisite excitations whose calculation procedure is as follows: (1) we choose $N_{ex}$ excited sites randomly according to excitation density. (2) we determine the excitation energy $E_{ex}$ of each site randomly according to a Gaussian distribution $G(E_m,\Delta E)=\exp[-\lbrace(E_{ex}-E_m)/\Delta E\rbrace^2]$, where $E_m$ and $\Delta E$ are the mean energy and the width, and are set to be 3.3 eV and 1.8 eV, respectively. These values correspond to the energy of the transient state of the photoexcited electron-hole pair spanning two graphite layers, and have been obtained in our previous paper.\cite{Nishioka} We assume that this randomness of the excitation energy originates from the uncertainty of the transient excited state. (3) we determine the initial potential energy $P$ and the kinetic energy $K$ randomly under the condition that $P+K=E_{ex}$. (4) we calculate the MD, starting from the initial state determined above, and count out the number of interlayer bonds formed after the time 1.0 ps. Within this time, the interlayer bond formation is expected to be almost complete. (5) we performed a similar calculation 50 times and average them.

\begin{figure}
\begin{center}
  \includegraphics[width=0.7\linewidth]{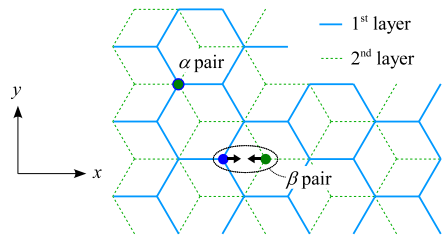}
  \caption{Schematic representation of $AB$ stacking of two graphite layers. The solid line indicates the first layer and the dashed line is the second one. The carbon pair in which two carbons locate at the same site in $xy$ plane is referenced as an $\alpha$ pair, and the pair which has a shift between two carbons in $x$ direction is referenced as a $\beta$ pair.}
  \label{fig:graphite}
\end{center}
\end{figure}

In the procedure (1), due to the $AB$ stacking, there are two types of carbon pairs, as shown in Fig. \ref{fig:graphite}. One is the case that the positions of two carbons at the first layer and at the second one are mutually same ($\alpha$ pair). Another is that the positions shift each other in $x$ direction ($\beta$ pair). In the latter case, the initial intrusion and velocity of the two excited carbons have not only the $z$ component but also the $x$ component, as indicated by arrows in Fig. \ref{fig:graphite}. Assuming only the $x$ polarized light irradiation, all the $\beta$ pairs have the shift only in the $x$ direction.

\section{results and discussions}\label{sec:results_discussion}

\subsection{cooperative phenomena}\label{subsec:cooperative_phenomena}

\begin{figure}
\begin{center}
  \includegraphics[width=1.0\linewidth]{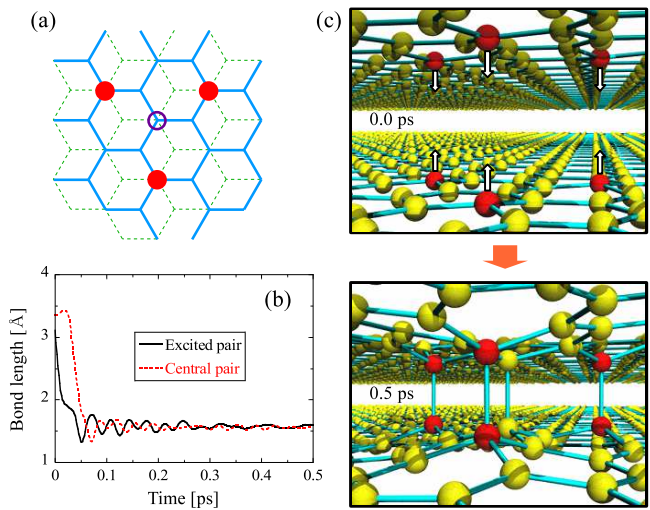}
  \caption{The example of cooperative phenomena in the case of 3 excited $alpha$ pairs with the energy of 4.22 eV/pair. (a) Schematic representation of two graphite layers. The $\alpha$ pairs indicated by red closed circle are excited initially, having the kinetic energy of 4.08 eV/pair and the potential energy of 0.14 eV/pair (the intrusion of 0.12 \AA). (b) The bond length between two opposite carbons as a function of time. The black solid line indicates the excite pair and the red dashed line is the central pair marked by a purple open circle in (a). (c) The snapshots of the MD calculation at time $t=$ 0 and 0.5 ps.}
  \label{fig:coopphen1}
\end{center}
\end{figure}

\begin{figure}
\begin{center}
  \includegraphics[width=1.0\linewidth]{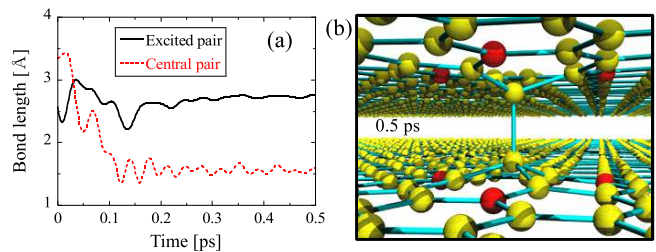}
  \caption{The example of cooperative phenomena in the case of 3 excited $alpha$ pairs with the energy of 2.51 eV/pair. The kinetic and potential energies are 1.04 eV/pair and 1.47 eV/pair (the intrusion of 0.39 \AA). (a) The bond length as a function of time, plotted in the same way as Fig.\ref{fig:coopphen1}(b). (b) The snapshots of the MD calculation at time $t=$ 0.5 ps.}
  \label{fig:coopphen2}
\end{center}
\end{figure}

\begin{figure}
\begin{center}
  \includegraphics[width=1.0\linewidth]{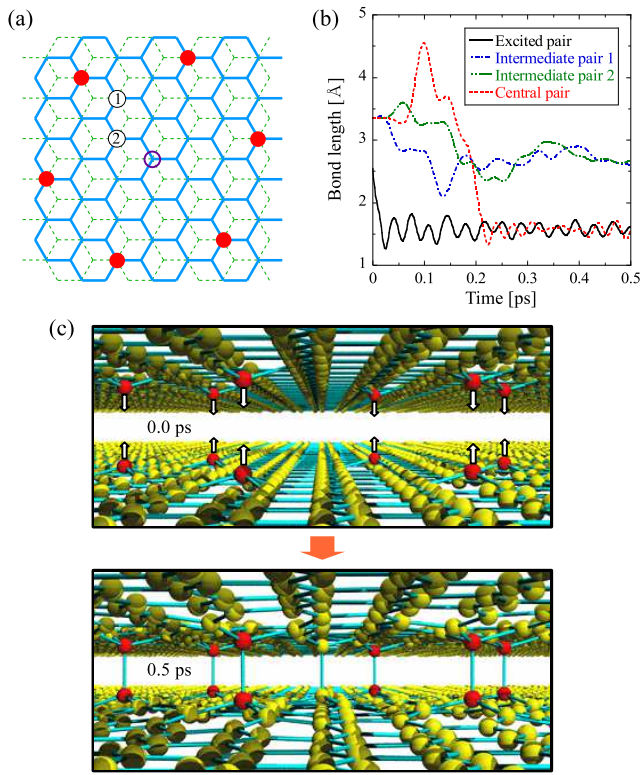}
  \caption{The example of cooperative phenomena in the case of 6 excited $alpha$ pairs with the energy of 4.74 eV/pair. The kinetic and potential energies are 3.20 eV/pair and 1.54 eV/pair (the intrusion of 0.4 \AA). (a) Schematic representation of two graphite layers, showing the initially excited $\alpha$ pairs indicated by red closed circle. (b) The bond length as a function of time, plotted in the same way as in the previous examples, while the one-dotted-dashed line (blue) and two-dotted-dashed line (green) denote the bond lengths of the intermediate $\alpha$ pairs numbered by 1 and 2 in (a). (c) The snapshots of the MD calculation at time $t=$ 0 and 0.5 ps.}
  \label{fig:coopphen3}
\end{center}
\end{figure}

We show the results for the cases where two or more carbon pairs are excited simultaneously. From the results of the calculations for various initial states, we have found various interesting cases. Here, we discuss three examples for such cases shown in Figures \ref{fig:coopphen1}, \ref{fig:coopphen2} and \ref{fig:coopphen3}. The first example is the case where three $\alpha$ pairs indicated by red closed circles in Fig. \ref{fig:coopphen1}(a) are excited. Each pair has the initial kinetic energy of 4.08 eV and the potential energy of 0.14 eV corresponding to the intrusion of 0.12 \AA. Hence, the total energy (4.22 eV/pair) is not enough to form an interlayer bond individually. We should note that the lowest energy of the bond formation is about 4.5 eV, as mentioned above. That is, no interlayer bond can be formed in this energy region without cooperative effects of multisite excitation.

Figure \ref{fig:coopphen1}(b) shows the time development of the bond length between two opposite carbons, where the black solid line indicates the originally excited pair and the red dashed line is the central pair marked by a purple open circle in Fig. \ref{fig:coopphen1}(a). As seen from this figure, the interlayer distance of the originally excited pair shrinks less than 2 \AA\, within 0.05 ps, resulting in an interlayer bond stable after 0.3 ps. With a little delay, even the interlayer distance of the central pair also shrinks, being dragged by the shrinkage of the bond length of the excited pairs, and results in a new interlayer bond.

We show the snapshots of the MD calculation at $t=$ 0 ps and 0.5 ps in Fig. \ref{fig:coopphen1}(c). In this example, due to the cooperation of the three closely excited pairs, three interlayer bonds can be formed at the originally excited sites, even though each excited pair does not have the sufficient energy to form the bond individually. Moreover, a new interlayer bond can be also formed at the central site, though it is not originally excited. This is because the extra energies from the originally excited pairs have propagated and concentrated at the center as a vibrational energy. This result means the nonlinearity that the number of interlayer bonds proliferates from 0 to 4, because of the cooperativity.

The second example is shown in Fig. \ref{fig:coopphen2}, where three $\alpha$ pairs are excited similar to the previous example, while the initial kinetic and potential energies are 1.04 eV/pair and 1.47 eV/pair corresponding to the intrusion of 0.39 \AA. Hence, the total energy is 2.51 eV/pair, being much lower than in the previous case. We also show the time development of the bond length and the snapshot at $t=$ 0.5 ps in Figures \ref{fig:coopphen2}(a) and (b), respectively. As seen from these figures, the bond length of the excited pair shrinks a little but does not reach up to 2 \AA. Of course, this is because the excitation energy is too small to form an interlayer bond at the excited sites. However, the energies of the excited pairs concentrate on the central site and are superposed. Therefore, the interlayer bond is formed there at about 0.1 ps, as the bond lenth of the central pair shows. This example also shows the nonlinearity that the number of interlayer bonds proliferates from 0 to 1.

Similarly, we show the third example in Fig. \ref{fig:coopphen3}, where six $\alpha$ pairs are excited. Compared with the previous examples, the number of initially excited sites is larger, but each excited site is more distant from others, instead. The excitation energy is 4.74 eV/pair, in which the kinetic and the potential energies are 3.20 eV/pair and 1.54 eV/pair (the intrusion of 0.4 \AA). In Fig. \ref{fig:coopphen3}(b), the one-dotted-dashed line (blue) and the two-dotted-dashed line (green) denote the bond lengths of the intermediate $\alpha$ pairs numbered by 1 and 2 in Fig. \ref{fig:coopphen3}(a), respectively. As seen from Fig. \ref{fig:coopphen3}(b), the bond length of the excited pair shrinks rapidly and an interlayer bond is formed there, since the original excitation energy is enough to form this bond. Furthermore, the extra energies of these excited pairs propagate as far as to the central site gradually, through the intermediate sites as a vibrational energy. Then, the bond length of the central pair expands once but afterwards shrinks rapidly, resulting in the interlayer bond at about 0.2 ps. The nonlinearity in this example indicates that the number of interlayer bonds proliferates from 6 to 7.

\begin{figure}
\begin{center}
  \includegraphics[width=0.7\linewidth]{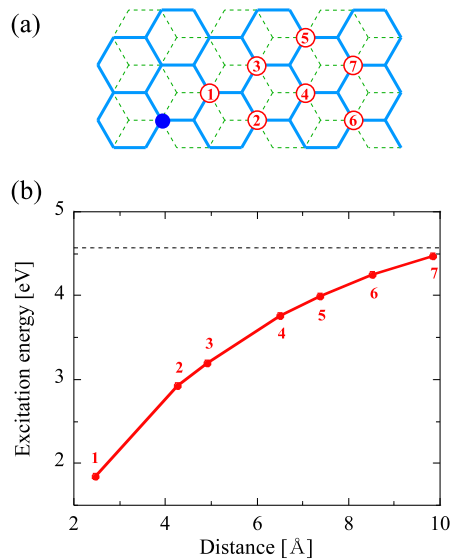}
  \caption{(a) The blue closed circle denotes the already existing interlayer bond and the excited sites$\alpha$ pairs are numbered from 1 to 7. (b) the lowest limit of the excitation energy as a function of the distance between the already existing bond and the excited site. The horizontal dashed line indicates the lowest limit energy to form an interlayer bond individually.}
  \label{fig:reqeng}
\end{center}
\end{figure}

Next, we discuss a more different cooperative phenomena from the previous ones. In the initial state of this case, an interlayer bond already exists at the site indicated by a blue closed circle in Fig. \ref{fig:reqeng}(a), and one of $\alpha$ pairs numbered from 1 to 7 by red is further excited. We performed the MD calculation starting from such an initial state for various excitation energies and obtain the lowest excitation energy necessary to form an interlayer bond. Figure \ref{fig:reqeng}(b) shows the lowest excitation energy as a function of the distance between the initially existing bond and the newly excited site. The numbering by red corresponds to that in Figure \ref{fig:reqeng}(a). The horizontal dashed line at about 4.5 eV in this figure means the individual lowest energy of interlayer bonding. As seen from this figure, the shorter the distance from the existing interlayer bond is, the smaller the necessary excitation energy becomes. That is, an interlayer bond is easier to be formed around the existing bond. This is because carbons around the interlayer bond are intruded inside of the layers. When the excited pair locates about 10 \AA\, away from the existing bond, the lowest energy is almost same as the case of the individual bond formation. These results also show the nonlinearity of the multisite excitation.

\subsection{random multisite excitation}\label{subsec:random_excitation}

\begin{figure}
\begin{center}
  \includegraphics[width=1.0\linewidth]{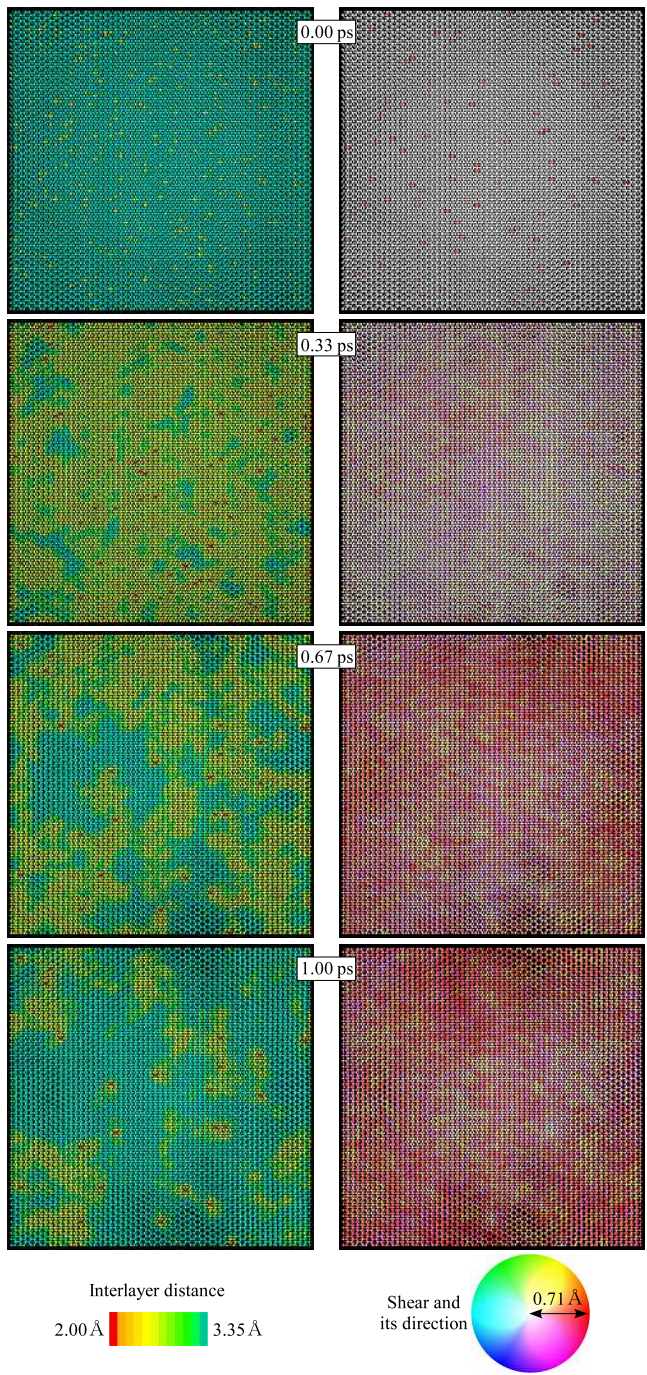}
  \caption{The snapshots of the MD calculation for the excitation density 5 \% at $t=$ 0, 0.33, 0.67 and 1.0 ps. The snapshots at the left side represent the distance between two graphite layers by the color palette at the left bottom, wherein the red means that an interlayer bond is formed. Those at the right side represent the shear of the upper layer to the lower layer and its direction by the circular color palette at the right bottom, wherein the white at the center of the circle means no shear and the shear of 0.71 \AA\, in the right direction is colored in red.}
  \label{fig:rex0.05}
\end{center}
\end{figure}

We have thus seen the several examples for cooperative phenomena. Next, we discuss the more realistic and random multisite excitation whose methodology has been already explained in detail in Sec. \ref{sec:model_method}. We calculate the random multisite excitation for several excitation densities and show the results. In Fig. \ref{fig:rex0.05}, as a example, we show the snapshots of the MD calculation for the excitation density 5 \%, that is, the number of excited pairs is 312, including $\alpha$ and $\beta$ pairs randomly. These snapshots are taken from the view over the graphite layers. Gazing at them, a lot of six-membered rings in the layers can be recognized. In the left figures, the distance between two graphite layers is represented by the color palette whose relation with the distance is plotted at the left bottom. Here, the red means that the distance becomes less than 2 \AA, that is, the interlayer bond is formed. In the right figures, the shear displacements of the upper layer relative to the lower one, together with its direction, are represented by the circular color palette plotted at the right bottom. The white at the center of the circle means no shear. For example, if the carbon of the upper layer shifts 0.71 \AA\, relatively to the carbon of the lower one in $x$ direction, these carbons are colored in red.

Let us see the snapshots at the left side of Fig. \ref{fig:rex0.05} in its time order. At $t=$ 0 ps, the excited pairs are colored in yellow or orange since they are given some intrusion to the inside of the layers initially. At 0.33 ps, due to the motion of the excited pairs to the inside, the shrinkage of the interlayer distance occurs in almost all the area and the interlayer bonds appear at some of the excited sites. Afterwards, the part of the layers which does not contain the interlayer bonds separates away toward the original interlayer distance, as seen from the snapshot at 0.67 ps. Finally, the system becomes almost stable with several interlayer bonds at 1.0 ps, wherein the shrinkage of the interlayer distance remains around the interlayer bonds. Of course, each carbon still keeps moving as a small vibration, since there is no energy dissipation in the system. In this 5 \% excitation case, as seen from the snapshot at 1.0 ps, the interlayer bonds, which finally remain, are rather rare. Therefore, the cooperative phenomena do not occur so intensively.

Next, let us see the snapshots of the shear at the right side of Fig. \ref{fig:rex0.05}. Since there is no shear at $t=$ 0 ps, the layers are white except the excited $\beta$ pairs. At a glance, they might not be recognized to be white. However, by gazing at the figure in detail, it can be seen that the carbons and the bonds are colored in white. When a $\beta$ pair is excited, the upper carbon shifts toward right relatively to the lower carbon, according to the initially given potential energy. Therefore, the excited $\beta$ pairs are colored in light red. As seen from the snapshots at 0.33 and 0.67 ps, most of the carbons in the layers becomes red gradually, that is, the shear toward the right direction occurs in all the area of the layers due to the motion of the excited $\beta$ pairs. It almost remains even at 1.0 ps. If two carbons of a $\beta$ pair are stably bonded, these carbons keep red color. However, any interlayer bonds are not actually formed in $\beta$ pairs. As a result of the MD calculation after 1.0 ps, it is confirmed that the shear disappears gradually, and all the area of the layers return to be white again at about 1.5 ps, though we do not show it here.

\begin{figure}
\begin{center}
  \includegraphics[width=1.0\linewidth]{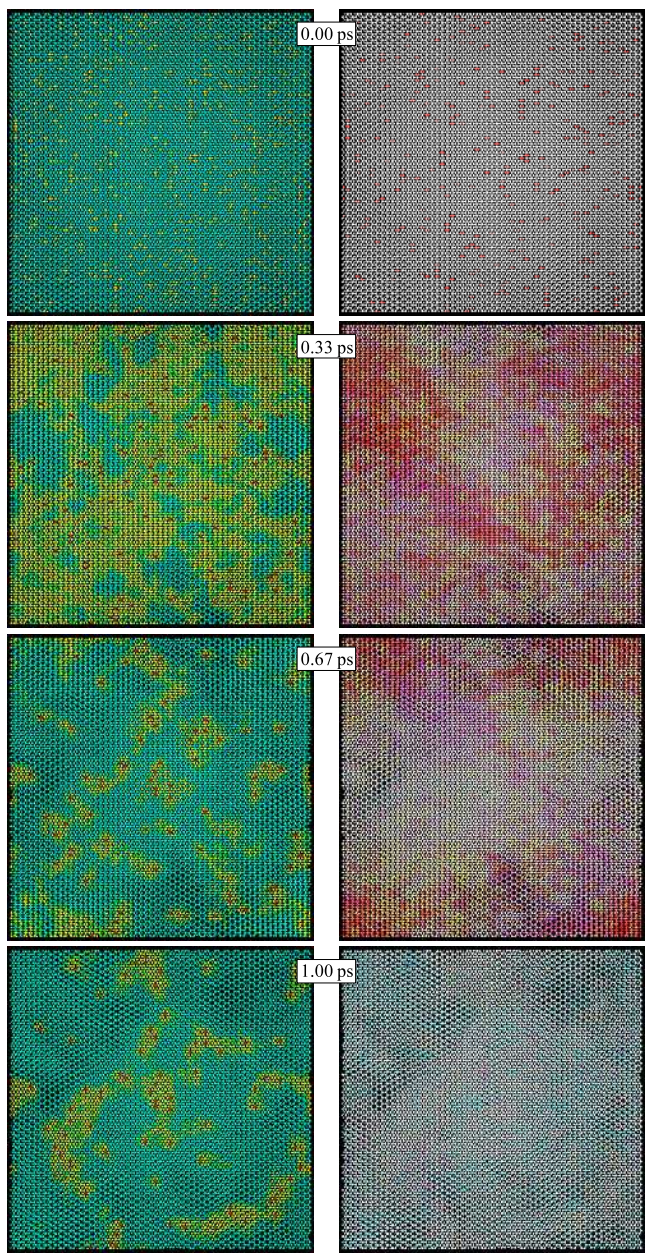}
  \caption{The snapshots of the MD calculation for the excitation density 10 \%, represented in the same way as Fig. \ref{fig:rex0.05}.}
  \label{fig:rex0.1}
\end{center}
\end{figure}

Similarly, as a example, we show the snapshots of the MD calculation for the excitation density 10 \%, corresponding to 624 pairs' excitation, in Fig. \ref{fig:rex0.1}. As seen from the left figures, the shrinkage of the interlayer distance occurs in almost all the area and the interlayer bonds appear at some of the excited sites, as well as in the case of 5 \% excitation. While this shrinkage appears faster and the number of the interlayer bonds is more abundant than in the previous case, because of the larger number of excitations. The shrinkage finally remains only around the interlayer bonds at 1.0 ps. Compared with the case of 5 \% excitation, the interlayer bonds cluster or close together. The cooperative phenomena occur much more intensively in this case.

As seen from the snapshots at the right side of Fig. \ref{fig:rex0.1}, the shear in the right direction also occurs in all the area once, but afterwards it disappears gradually. The occurrence of the shear and its disappearance are also faster than in the 5 \% excitation case. Moreover, the layers shift in the opposite direction only a little, as shown by a lot of carbons colored in light blue at 1.0 ps. This means that no interlayer bond is formed in $\beta$ pairs.

\begin{figure}
\begin{center}
  \includegraphics[width=0.9\linewidth]{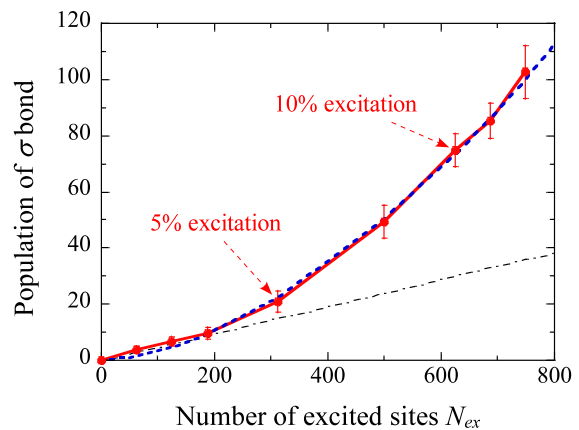}
  \caption{The number of interlayer $\sigma$ bonds as a function of the number of excited sites $N_{ex}$. The black dotted dashed line indicates the case where the interlayer bond increases proportionally to $N_{ex}$, whose slope is 0.048. The blue dashed line is a fitted curve proportional to $(N_{ex})^{1.7}$.}
  \label{fig:nonlinear}
\end{center}
\end{figure}

In Fig. \ref{fig:nonlinear}, we show the number of final interlayer bonds, counted at 1.0 ps and averaged over 50 calculations, as a function of the number of excited sites $N_{ex}$. It is plotted by the red solid line. The dotted dashed line (black) indicates the case where the interlayer bonds would increase linearly proportional to $N_{ex}$. This slope is obtained as the probability that an interlayer bond is formed when one carbon pair alone is excited randomly. As seen from this figure, the number of interlayer bonds for more than 5 \% excitation increases nonlinearly with respect to the number of excited sites, and can be fitted by the blue dashed curve $\propto(N_{ex})^{1.7}$. This means that effectively 1.7 excitations contribute to the interlayer bond formation. While there is no cooperative phenomenon at all for less than 3 \% excitation. In our calculation, the mean energy of the excitation is 3.3 eV, corresponding to about 2.1 photons with the energy of 1.57 eV. That is, 1.7 excitations is equivalent to $1.7\times2.1=3.57$ photons on average. Therefore, we conclude that about three or four photons are effectively involved in this process.

Incidentally, we discuss the shear between graphite layers. The existence of the shear is very important for the diaphite domain formation, as discussed in the previous works,\cite{Kanasaki,Radosinski1,Ohnishi1,Ohnishi2,Radosinski2} since interlayer bonds in $\beta$ pairs are necessary to build up the diaphite structure. In our present calculation, the shear once appears but disappears as time passes. We think it is because the excitation in this case is a pulse, or only at the very beginning. As long as the light is irradiated for a certain period of time, the shear can be induced by the fast part of this period, and also the subsequent excitation can accelerate this shear by the later part of the period before it disappears. Therefore, it is probable that interlayer bonds are successfully formed in $\beta$ pairs, so that the shear remains stably, resulting in the diaphite domain. The calculation which takes into account the time width of the excitation is our future work.

\section{summary}\label{sec:summary}
We have thus investigated the nonlinearity in the interlayer $\sigma$ bond formation, using the classical molecular dynamics and the Brenner's potential. Due to the multisite excitation, various cooperative phenomena between excited sites appear, so that the interlayer bonds, more than the number of excited sites, are formed, or making up for the lack of energy each other. In the situation where an interlayer bond already exists, it is easier to form a new interlayer bond nearby. Based on the understanding of such cooperative phenomena, we have studied the random multisite excitation. As a result, the number of formed interlayer bonds increases nonlinearly with respect to the number of excited sites. The nonlinearity shows 1.7 power of the number of excited sites. This result means that about three or four photons are effectively involved in the interlayer bond formation.

\section{acknowledgments}
The authors thank K. Tanimura, J. Kanasaki, E. Inami, and H. Ohnishi for presenting their results prior to publication and valuable discussions. This work is supported by the Ministry of Education, Culture, Sports, Science and Technology of Japan, the peta-computing project, and Grant-in-Aid for Scientific Research (S), Contract No. 19001002, 2007.

\end{document}